\documentclass[12pt]{article}
\usepackage{epsfig,amsfonts,amsthm}
\usepackage{amsmath,amssymb}
\newcommand{\be}{\begin{equation}}
\newcommand{\ee}{\end{equation}}
\newcommand{\bea}{\begin{eqnarray}}
\newcommand{\eea}{\end{eqnarray}}

\newcommand{\f}{\phi}
\newcommand{\fd}{\phi^\dagger}
\renewcommand{\Re}{\mathrm{Re }}
\renewcommand{\Im}{\mathrm{Im }}

\newcommand{\doublet}[2]{ \left( \begin{array}{c}#1 \\ #2 \end{array}\right) }

\newcommand{\lr}[1]{ \langle #1 \rangle}

\def\lsim{\mathrel{\rlap{\lower4pt\hbox{\hskip1pt$\sim$}}
    \raise1pt\hbox{$<$}}}         
\def\gsim{\mathrel{\rlap{\lower4pt\hbox{\hskip1pt$\sim$}}
    \raise1pt\hbox{$>$}}}         

\title{Frustrated symmetries in multi-Higgs-doublet models}

\author{Igor~P.~Ivanov$^{1,2}$, Venus~Keus$^{1}$
\\
  {\small $^1$ IFPA, Universit\'{e} de Li\`{e}ge, All\'{e}e du 6 Ao\^{u}t 17, b\^{a}timent B5a, 4000 Li\`{e}ge, Belgium}\\
  {\small $^2$ Sobolev Institute of Mathematics, Koptyug avenue 4, 630090, Novosibirsk, Russia}\\
  }

\begin{document}
\maketitle

\begin{abstract}
Within multi-Higgs-doublet models, one can impose symmetries on the Higgs potential, either discrete or continuous,
that mix several doublets. In two-Higgs-doublet model any such symmetry can be conserved
or spontaneously violated after the electroweak symmetry breaking (EWSB), depending on the 
coefficients of the potential. With more than two doublets, there exist symmetries
which are always spontaneously violated after EWSB.
We discuss the origin of this phenomenon and show its similarity to
geometric frustration in condensed-matter physics.
\end{abstract}

\section{Introduction}

The electroweak theory relies on the Higgs mechanism of the electroweak symmetry breaking (EWSB). 
Many different variants of EWSB beyond the Standard model have been proposed so far, see e.g. \cite{CPNSh},
the $N$-Higgs-doublet models (NHDM) being among the most conservative ones. 
On one hand, various realizations of NHDM are attractive to a phenomenologist
because they offer a broad spectrum of new physics phenomena with little input.
On the other hand, they naturally arise in supersymmetric models and in certain low-energy 
realizations of superstring/brane models, see \cite{gupta2010} and references therein.
Thus, the multi-doublet models clearly deserve a detailed study from different points of view.

When building a multi-Higgs-doublet model, one has to specify the Higgs potential.
The number of free parameters in the potential explodes as $N$ increases.
Even for the two-Higgs-doublet model (2HDM) one already has 14 free parameters,
which makes the minimization of the potential intractable with the straightforward algebra
and necessitates introduction of more involved mathematical methods to study
properties of a sufficiently general 2HDM, see \cite{2HDM} and references therein. 
Generalization of these methods to general $N$ is a hard task; the first steps were recently
made in \cite{NHDM2010,NHDM2010potential}.

One of the attractive features of NHDM is the possibility to introduce additional symmetries
in the space of Higgs families via the Higgs potential, 
which can then affect the other sectors of the theory.
In 2HDM, the full list of symmetries of the potential is now known \cite{2HDM}: the symmetry group
can be $Z_2$, $(Z_2)^2$, $(Z_2)^3$, $O(2)$, $O(2)\times Z_2$, or $O(3)$.
Trying to extend these symmetries into the Yukawa sector, one can encounter models with
remarkable $CP$-properties, \cite{remarkableCP}.
With more than two doublets, classifying possible symmetries of the potential 
seems to be a rather complicated task. Just to mention several recent attempts to attack the problem,
in \cite{abelian3HDM} the Abelian symmetries of 3HDM were classified, 
in \cite{ludl} many group-theoretic aspects of the Yukawa mass matrices in 3HDM were analyzed,
in \cite{Olaussen2010} some symmetry features of multi-doublet models and their traces 
in the Higgs mass spectrum were presented, while \cite{NHDM2010potential} contains
a general strategy to identify discrete symmetry groups of NHDM, with several 3HDM examples.
However, the full list of symmetries that can be imposed on the scalar potential of 
a multi-Higgs-doublet model is still unknown.

In this paper we would like to draw attention to a peculiar feature that arises in models with more
than two Higgs doublets. We show that there exist certain explicit symmetries of the potential,
which are necessarily broken whenever the electroweak symmetry breaking takes place.
We term them ``frustrated symmetries'' because of their similarity to the phenomenon of
geometric frustration in condensed-matter physics.
We think that such symmetries can be an interesting option when building models 
with desired properties beyond the Standard model.

The structure of the paper is the following. 
In Section \ref{section2} we introduce a convenient framework for discussing symmetries of the Higgs potential.
Then in Section \ref{section3} we explain how frustrated symmetries arise and give 3HDM examples.
In Section \ref{section4} we make several remarks and draw our conclusions.

\section{Orbit space of NHDM}\label{section2}

A detailed description of the orbit space of NHDM was given in \cite{NHDM2010}.
Here we just introduce the notation and mention some of its results.

In the $N$-Higgs-doublet model we introduce $N$ Higgs doublets $\phi_a$, $a=1,\dots,N$.
The general renormalizable Higgs potential of NHDM is constructed from the gauge-invariant
combinations $(\phi_a^\dagger \phi_b)$, which describe the gauge orbits in the Higgs space. 
The space of gauge orbits (the orbit space) can be represented as a certain algebraic manifold
embedded in the euclidean space $\mathbb{R}^{N^2}$ of bilinears
\be
r_0 = \sqrt{{N-1\over 2N}}\sum_a \phi_a^\dagger \phi_a\,,\quad r_i = \sum_{a,b} \phi_a^\dagger \lambda^i_{ab}\phi_b\,,
\label{rmu}
\ee
where $\lambda^i$ are the generators of $SU(N)$: Pauli matrices for $N=2$, Gell-Mann matrices for $N=3$, etc.
The Higgs potential can then be rewritten as a quadratic polynomial in bilinears $r_0$ and $r_i$:
\be
\label{potential}
V = - M_0 r_0 - M_i r_i + {1 \over 2}\Lambda_{00} r_0^2 + \Lambda_{0i} r_0 r_i + {1 \over 2}\Lambda_{ij} r_i r_j\,.
\ee
Note that when constructing a potential, we have the full freedom to choose $N^2$ components $M_0$, $M_i$ 
and $N^2(N^2+1)/2$ components of $\Lambda_{00}$, $\Lambda_{0i}$, and $\Lambda_{ij}$, provided 
the positivity constraints are satisfied\footnote{The explicit algebraic formulation of the positivity
constraints in the most general NHDM is not yet known, 
however it is easy to satisfy them when constructing potentials with specific symmetries, see \cite{NHDM2010potential}.}.

Some geometric properties of the orbit space manifold were established in \cite{NHDM2010}.
It is located inside a conical ``shell'' lying between two coaxial cones, the ``forward lightcone'' and a certain inner cone:
\be
{N-2 \over 2(N-1)} \le \vec n^2 \le 1\,,\quad n_i \equiv {r_i \over r_0}\,.
\ee
If the vector $\lr{\vec n}$ corresponding to the vacuum expectation values lies on the surface of the ``forward lightcone'',
$\lr{\vec n^2} = 1$, then the vacuum is neutral; otherwise, the vacuum is charge-breaking.
In the case of two-Higgs-doublet model, $N=2$, the inner cone disappears, and the orbit space
populates the entire unit ball in the 3D $\lr{\vec n}$-space, including the point $\lr{\vec n} =0$,
which corresponds to the following v.e.v.'s:
\be
\lr{\phi_1} =\frac{1}{\sqrt{2}} \doublet{0}{v}\,,\quad \lr{\phi_2} =\frac{1}{\sqrt{2}} \doublet{v}{0}\,.\label{maximal2HDM}
\ee
The inner cone appears for $N>2$ and makes the innermost part of the unit ball not realizable in terms of doublets.
For example, for three-Higgs-doublet model, $1/2 \le |\vec n| \le 1$.

Note also that if we were working with two Higgs singlets, or if we were working in 2HDM but required that the minima
always be neutral, then the orbit space available would be represented by a unit sphere, not ball.
In this case, the point $\lr{\vec n} =0$ would not be realizable anymore.

\section{Frustrated symmetries}\label{section3}

\subsection{General observations}

Suppose we have constructed a Higgs potential with an explicit symmetry.
That is, we have found coefficients $M_i$, $\Lambda_{0i}$ and $\Lambda_{ij}$ in
(\ref{potential}), such that the potential is left invariant under a group $G$ of 
transformation between doublets $\phi_a$. We will restrict ourselves only to unitary or antiunitary transformations,
which leave the Higgs kinetic term invariant. This means that $r_0$ is invariant under $G$, 
while the components $r_i$ are transformed by orthogonal transformations forming a subgroup
of $O(N^2-1)$.

The key observation is that imposing a symmetry often restricts $M_i$ and $\Lambda_{0i}$ stronger than $\Lambda_{ij}$.
In particular, it is possible to devise such a symmetry group $G$ of transformations of doublets, 
which can be implemented via a non-trivial tensor  $\Lambda^{(G)}_{ij}$,
but not in a vector $M_i$ or $\Lambda_{0i}$ (see examples below). The Higgs potential with such a symmetry is then
\be
\label{potential2}
V^{(G)} = - M_0 r_0 + {1 \over 2}\Lambda_{00} r_0^2 + {1 \over 2}\Lambda^{(G)}_{ij} r_i r_j\,.
\ee
The group-theoretic explanation of this possibility comes from counting singlets in different representations.
The vector $r_i$ realizes the adjoint representation of the group $G$, while $r_ir_j$ transforms as
a product of two such representations.
When constructing a $G$-invariant potential, we look for singlets in $r_i$ or in $r_ir_j$.
One can easily think of situations when there is no singlet in $r_i$, while there exists a singlet in $r_ir_j$.
For such a group, the Higgs potential has no linear $r_i$ term, 
but can contain various $\{r_ir_j\}_{singlet}$ terms.

After EWSB, the doublets acquire some vacuum expectation values, which are translated into certain 
$\lr{r_0} > 0$ and $\lr{r_i}$. If it turns out that all $\lr{r_i} = 0$, the vacuum remains invariant
under the symmetry group $G$. Otherwise, if at least one component $\lr{r_i} \not = 0$, then
the vacuum spontaneously breaks the symmetry group $G$, either completely or down to a proper subgroup,
simply because by construction no non-trivial vector $\lr{r_i}$ can be $G$-symmetric.
Thus, in order to see if a given symmetry has a chance to be conserved in a non-trivial vacuum,
we need to check whether or not the point $\vec n = \lr{\vec r}/\lr{r_0} = 0$ belongs to the orbit space of the model.

In the previous section we showed that the point $\vec n = 0$ belongs to the orbit space 
only in the two-Higgs-doublet model. Therefore, in 2HDM any symmetry imposed on the potential can, in principle,
be conserved in the vacuum state provided we have chosen appropriate coefficients of the potential.
Indeed, if one acts on the state (\ref{maximal2HDM}) with any (anti)unitary transformation which mixes the doublets,
one arrives at the same state up to an electroweak transformation. That is, the corresponding gauge orbit
is $G$-symmetric.

In NHDM with $N>2$, the point $\vec n = 0$ cannot be realized through doublets; 
therefore, the minimum of the $G$-symmetric potential (\ref{potential2}) unavoidably breaks the explicit symmetry
in the space of electroweak orbits. Spontaneous breaking of this symmetry always accompanies EWSB.

We call such a symmetry a {\bf frustrated symmetry} for the following reason:
setting all $n_i = 0$ is equivalent to setting all products $(\fd_i \f_j)$ 
to zero while keeping all the norms $|\phi_i|^2$ non-zero and equal.
This is impossible to achieve with more than two doublets, 
simply because we have too little freedom of where
to place vacuum expectation values inside all three doublets. 
Even if the first two doublets are chosen as in (\ref{maximal2HDM}),
the third doublet of the same norm will have a non-zero product with the first or the second doublet.
In other words, although we can ``optimize'' the v.e.v.'s in any pair of doublets (equal norms, zero product),
these optimal configurations are mutually incompatible in three or more doublets and cannot be satisfied simultaneously.
This is precisely what geometric frustration in condensed-matter physics is about.
 
\subsection{3HDM examples}

Let us now give three examples of frustrated symmetries in the three-Higgs-doublet model (3HDM).

The simplest case is given by the potential which is symmetric
under any $SU(3)$ rotation and mixes the three doublets. It corresponds to the exceptional case $\Lambda_{ij} =0$:
\be
V = - m^2 (\fd_1 \f_1 + \fd_2 \f_2 + \fd_3 \f_3) + \lambda (\fd_1 \f_1 + \fd_2 \f_2 + \fd_3 \f_3)^2\,.
\ee 
Clearly, no non-trivial vacuum can be symmetric under the entire $SU(3)$ group, even up to EW transformations.
However, this example involves a continuous group, whose breaking generates undesired massless scalars after EWSB. 

To avoid massless scalars, one can impose discrete frustrated symmetries.
Our second example is given by the ``tetrahedral 3HDM'' defined by the following Higgs potential:
\bea
V & = &  - M_0 r_0 + \Lambda_0 r_0^2 + \Lambda_1(r_1^2+r_4^2+r_6^2) + \Lambda_2(r_2^2+r_5^2+r_7^2) 
+ \Lambda_3(r_3^2+r_8^2)\nonumber\\
&& + \Lambda_4(r_1r_2 - r_4r_5 + r_6r_7)\,. \label{tetrahedral3HDM}
\eea
In terms of doublets, this potential has the form:
\bea
V&=& - {M_0 \over \sqrt{3}} (\fd_1 \f_1 + \fd_2 \f_2 + \fd_3 \f_3) 
+ {\Lambda_0 +\Lambda_3 \over 3} \left[(\fd_1 \f_1)^2 + (\fd_2 \f_2)^2 + (\fd_3 \f_3)^2\right] \nonumber\\[2mm]
&&+ {2\Lambda_0 -\Lambda_3 \over 3} 
\left[(\fd_1 \f_1)(\fd_2 \f_2) + (\fd_2 \f_2)(\fd_3 \f_3) + (\fd_3 \f_3)(\fd_1 \f_1)\right]\nonumber\\[2mm]
&&+ \Lambda_1 \left[(\Re \fd_1\f_2)^2 + (\Re \fd_2\f_3)^2 + (\Re \fd_3\f_1)^2\right] \nonumber\\
&&+ \Lambda_2 \left[(\Im \fd_1\f_2)^2 + (\Im \fd_2\f_3)^2 + (\Im \fd_3\f_1)^2\right] \nonumber\\
&&+ \Lambda_4 \left[(\Re \fd_1\f_2) (\Im \fd_1\f_2) + (\Re \fd_2\f_3) (\Im \fd_2\f_3) + (\Re \fd_3\f_1) (\Im \fd_3\f_1)\right]\,.
\label{tetrahedral3HDMfields}
\eea
This potential is symmetric under the full (achiral) tetrahedral symmetry group $T_d$, which is isomorphic to $S_4$. 
Its elements are independent sign flips of the doublets, the cyclic permutation of the three doublets 
$\phi_1 \to \phi_2 \to \phi_3 \to \phi_1$, antiunitary transformations involving exchange of a pair of doublets 
applied together with the $CP$-transformation such as $\phi_1 \leftrightarrow \phi_2^\dagger$, $\phi_3 \to \phi_3^\dagger$, 
as well as their combinations.
Note the absence of any term which is linear in $r_i$ in (\ref{tetrahedral3HDM}) due to the fact that 
no such term respects the tetrahedral symmetry. 

If all $\Lambda$'s except for $\Lambda_1$ are positive and if $\Lambda_4/\sqrt{12} < |\Lambda_1| < \Lambda_0$, 
then the potential has four neutral degenerate global minima at
\be
\lr{\phi_i^0} = {v \over \sqrt{2}}\left\{ 1,\, \pm 1,\, \pm 1 \right\}\,,\quad v^2 = {M_0 \over \sqrt{3}(\Lambda_0 - |\Lambda_1|)}\,.
\label{vevtetrahedral3HDM}
\ee
Using the geometric technique developed in \cite{NHDM2010potential}, one can show that these are the global minima 
and they correspond to the vertices of a regular tetrahedron in the $\{n_1,\,n_4,\,n_6\}$-space at $n_2=n_3=n_5=n_7=n_8=0$, 
which are the contact points of the critical equipotential surface with the orbit space of 3HDM. 
At each of these minima, the tetrahedral symmetry group is broken down to a $D_3$ group,
the symmetry group of the equilateral triangle. The ``broken'' transformations link 
the minima to each other.

The third 3HDM example of frustrated symmetry is the ``octahedral 3HDM'', whose potential
is again given by (\ref{tetrahedral3HDM}) but with $\Lambda_4 = 0$. 
The resulting potential is symmetric under the octahedral group,
which in addition to the tetrahedral group of transformations includes 
the exchanges of the doublets, e.g. $\phi_1 \leftrightarrow \phi_2$, and the
$CP$-transformation $\phi_i \to \phi_i^\dagger$, separately, as well as their compositions with other transformations. 
If $\Lambda_1 < 0$ as before, the v.e.v.'s (\ref{vevtetrahedral3HDM}) again realize the global minima of this potential.

We note in passing a remarkable phenomenological feature of this model: it is 2HDM-like.
Due to remaining symmetry, it exhibits certain degeneracy in the mass spectrum of the physical Higgs bosons, 
yielding just one mass for both charged Higgs bosons and three different masses for the neutral ones,
which precisely mimics the typical Higgs spectrum of 2HDM.
It would be interesting to see what experimental observables could distinguish this model from the actual 2HDM.

Finally, we note that the octahedral symmetry can be implemented in a slightly different fashion described
in \cite{NHDM2010potential}. In that case, there are six degenerate vacua with symmetry group of a square.

\section{Discussion and conclusions}\label{section4}

Let us make several additional remarks concerning the frustrated symmetries and put forth questions
that deserve closer study.

First, frustrated symmetries are not specific to doublets.
They can arise when the representation of the electroweak group has lower dimensionality than
 the horizontal (Higgs family) space, i.e. more than one singlet, 
more than two doublets, more than three triplets, etc. 

Second, various cyclic groups often used in constructing
multi-Higgs-doublet models do not represent a frustrated symmetry.
Indeed, a cyclic group can be realized as a group of phase rotations
of the individual doublets. But these phase rotations leave the $N-1$ coordinates $(r_i)$ invariant,
corresponding to generators $\lambda_i$ of the Cartan subalgerba of $su(N)$.
Thus, the vacuum state with $\lr{\phi_1} \not = 0$ and all other $\lr{\phi_i}=0$ is invariant under this group
(up to an overall phase rotation). One can then easily construct a potential with the cyclic symmetry
whose global minimum will be exactly at that point and will, therefore, conserve the symmetry of the potential.

One might wonder whether frustrated symmetries must be non-Abelian. 
We have at least one counterexample to this conjecture: in a model with two singlets, 
the $(Z_2)^3$ symmetry is frustrated, see remark at the end of Section \ref{section2}.

Third, any non-trivial Yukawa sector of the model obviously violates a frustrated symmetry.
However, if the reparametrization transformation is generalized to include simultaneous 
transformations of scalar and fermionic fields, then there is a chance to extend a frustrated symmetry to the full
Lagrangian. Whether or not any frustrated symmetry can be extended to the fermion sector in this manner, remains to be studied.

Last, looking at the problem from a more phenomenological side, it should be noted that 
a frustrated symmetry is not manifested in the properties of physical Higgs bosons after EWSB.
Indeed, the vacuum expectation values and the Higgs mass spectrum necessarily violate the symmetry.
Still, knowing that the Higgs potential has a symmetry before EWSB is important because
it might provide hints of the origin of the Higgs families. The following question then arises:
is it possible to infer its presence from the experimental data,
and if so, what observables should one look at? 
A detailed phenomenological study is needed to clarify this issue.
\\

In conclusion, we showed that in multi-Higgs-doublet models with more than two doublets
one can impose symmetries on the Higgs potential which are necessarily broken after EWSB.
We named them frustrated symmetries because of their resemblance to the phenomenon of geometric frustration 
in condensed matter physics.
We discussed the group-theoretic and geometric origin of such symmetries and gave several examples
in the three-Higgs-doublet model.

\section*{Acknowledgements} 
Useful comments by J.-R. Cudell are acknowledged.
This work was supported by the Belgian Fund F.R.S.-FNRS via the
contract of Charg\'e de recherches, and in part by grants
RFBR No.08-02-00334-a and NSh-3810.2010.2.

\end{document}